\documentclass[apjl,tighten]{emulateapj}
\usepackage{epsfig}
\usepackage{amsmath,amssymb,bm}
\usepackage{color}
\usepackage[varg]{txfonts}

\begin{document}
\label{firstpage}

\title[Giant Outbursts in Be/X-ray Binaries]{Giant Outbursts in
  Be/X-ray Binaries}

\author{Rebecca G. Martin\altaffilmark{1,4}} \author{Chris
  Nixon\altaffilmark{1,5}} \author{Philip J. Armitage\altaffilmark{1}}
\author{Stephen H. Lubow\altaffilmark{2}} \author{Daniel
  J. Price\altaffilmark{3}} \affil{\altaffilmark{1}JILA, University of
  Colorado \& NIST, UCB 440, Boulder, CO 80309, USA}
\affil{\altaffilmark{2}Space Telescope Science Institute, 3700 San
  Martin Drive, Baltimore, MD 21218, USA }
\affil{\altaffilmark{3}Monash Centre for Astrophysics (MoCA), School
  of Mathematical Sciences, Monash University, Vic. 3800, Australia}
\affil{\altaffilmark{4}Sagan Fellow} \affil{\altaffilmark{5}Einstein
  Fellow}

\begin{abstract}
Be/X-ray binary systems exhibit both periodic (Type~I) X-ray outbursts
and giant (Type~II) outbursts, whose origin has remained elusive.  We
suggest that Type~II X-ray outbursts occur when a highly misaligned
decretion disk around the Be star becomes eccentric, allowing the
compact object companion to capture a large amount of material at
periastron.  Using 3D smoothed particle hydrodynamics simulations we
model the long term evolution of a representative Be/X-ray binary
system.  We find that periodic (Type~I) X-ray outbursts occur when the
neutron star is close to periastron for all disk inclinations. Type~II
outbursts occur for large misalignment angles and are associated with
eccentricity growth that occurs on a timescale of about 10 orbital
periods. Mass capture from the eccentric decretion disk results in an
accretion disk around the neutron star whose estimated viscous time is
long enough to explain the extended duration of Type~II
outbursts. Previous studies suggested that the outbursts are caused by
a warped disk but our results suggest that this is not sufficient, the
disk must be both highly misaligned and eccentric to initiate a
Type~II accretion event.
\end{abstract}

\section{Introduction}
\label{intro}

Be stars are early main-sequence stars which have shown H$\alpha$
emission at least once \citep[e.g.][]{Porter2003}. They are rapidly
rotating close to their break up velocity
\citep{Slettebak1982,Porter1996}. The stars are surrounded by
decretion disks \citep{Pringle1991}, formed from ejected material that
settles into a Keplerian disk and spreads outwards through viscous
diffusion \citep{Lee1991,Hanuschik1996,Carciofi2011}. The mechanism
for ejection is uncertain (rotation alone is insufficient), with
magnetic fields \citep[e.g.][]{Underhill1987,Smith1994} or stellar
pulsation \citep{Rivinius2001} being candidates.  The disk and its
associated emission appears and disappears on a timescale of a few
months to a few years \citep[e.g.][]{Bjorkman2002}.

Gas from the {\em decretion} disk of the Be star can be captured and
{\em accreted} by a companion in Be/X-ray binaries. These systems
contain a Be star in a relatively wide orbit (orbital period of tens
to hundreds of days) of significant eccentricity ($e\gtrsim 0.1$) with
a compact object, often a neutron star \citep{Reig2011} or less
commonly a black hole \citep{MunarAdrover2014}. Typically the orbital
angular momentum of the binary is misaligned to the spin of the Be
star, with a relative inclination of $i \gtrsim 25^\circ$ (see Table 1
in \citealt{Martinetal2011} and references therein). Both the
misalignment and the orbital eccentricity are plausibly the result of
kicks imparted during the supernova that formed the compact object
\citep[e.g.][]{Martinetal2009b,Martinetal2010}.

In this paper, we investigate how the interaction of the misaligned
disk and eccentric binary may give rise to the observed X-ray
phenomenology of Be/X-ray binaries.  Two types of X-ray outbursts are
produced in such systems \citep{Stella1986,Negueruela1998}. {\it Type
  I} X-ray outbursts are periodic and coincide with periastron passage
with a typical luminosity of $L_{\rm X} = 10^{36}-10^{37}~{\rm
  erg~s}^{-1}$ that lasts for a few days. {\it Type II} X-ray
outbursts occur much less frequently but have a larger luminosity of
$L_{\rm X} > 10^{37}~{\rm erg~s}^{-1}$ and last weeks to months,
showing no orbital modulation. The majority of giant outbursts are
finished in less than an orbital period, but some may last up to five
orbital periods \citep{Kretschmar2013}. Several suggestions as to the
origin of these outbursts have been made
\citep{Negueruela2001,Okazaki2001,Moritani2013,Okazaki2013}, all of
which involve as a central element the tilted and warped disks that
have been observed before and during the outbursts
\citep{Negueruela2001,Reigetal2007,Moritanietal2011}.  Most recently,
\cite{Kato2014}, presented analytic arguments suggesting that
simultaneous growth of eccentric and tilt modes could occur on
time-scales relevant to the giant outbursts. Exactly how a warped (and
potentially eccentric) disk causes a long duration accretion event
onto the neutron star remains, however, unclear.

The outline of the paper is as follows. In \S2 we present results from
three dimensional numerical simulations of accretion in Be/X-ray
binaries.  Type~I outbursts can be produced when the neutron star
captures a small fraction of low angular momentum gas from the disk at
closest approach. We show that Type~II outbursts can be explained as a
consequence of eccentricity growth within the Be star disc.
Capture of larger amounts of gas from the eccentric disk generates a
Type~II outburst, and in \S3 we show that the long duration of these
events can be due to the increased viscous time when the material
circularizes around the compact object.  We discuss our results in \S4
and \S5.

\section{Decretion disk around the Be Star}

\begin{figure*}
\includegraphics[width=18cm]{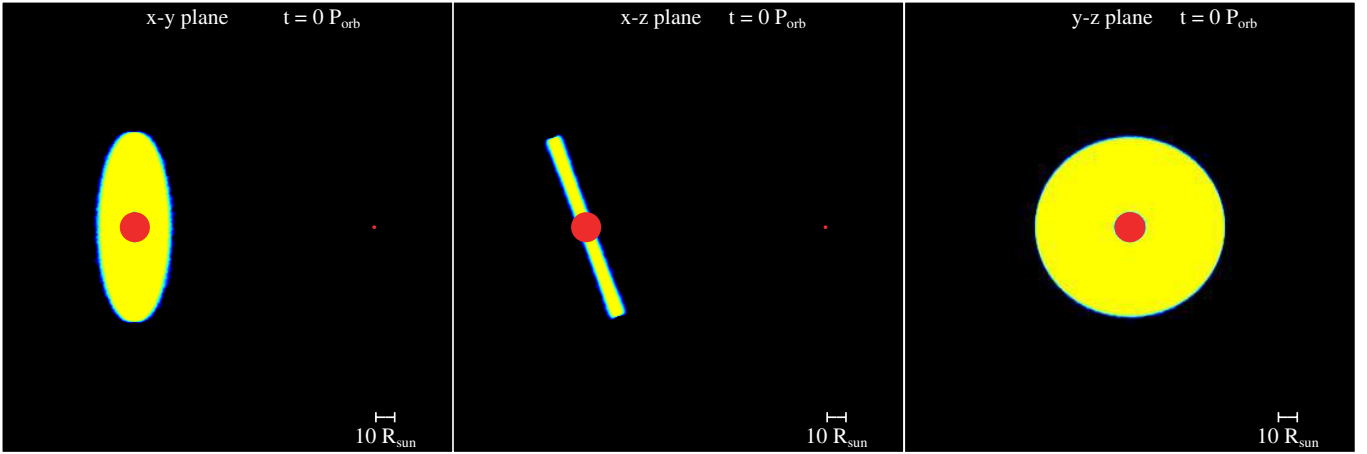}
\caption{Initial conditions for the SPH simulation of a Be star (large
  red circle) with a disk and a neutron star (small red circle) binary
  companion. The size of the circles denotes their respective
  accretion radii. The colour of the gas denotes the column density
  with yellow being about two orders of magnitude larger than blue.
  The left panel shows the view looking down on to the $x$-$y$ binary
  orbital plane and the middle and right panels show the views in the
  binary orbital plane, the $x$-$z$ and $y$-$z$ planes, respectively.
  Initially the disk is circular and flat, but tilted from the binary
  orbital plane by $70^\circ$. Note that in the right hand panel the
  neutron star and the Be star coincide with each other. }
\label{plot1}
\end{figure*}

\begin{figure*}
\begin{center}
\includegraphics[width=18cm]{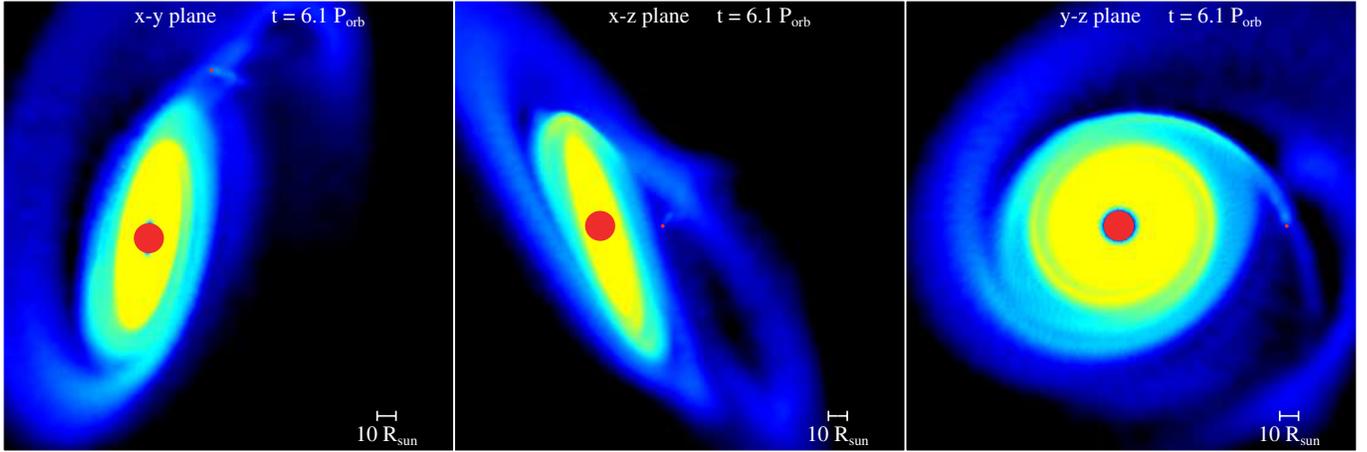}
\end{center}
\caption{Same as Fig.~\ref{plot1} but at a time of $6.1\,P_{\rm. orb}$.
  The Be star disk is close to circular and the neutron star is moving
  towards periastron and undergoing a Type~I outburst. }
\label{plot2}
\end{figure*}

\begin{figure*}
\begin{center}
\includegraphics[width=18cm]{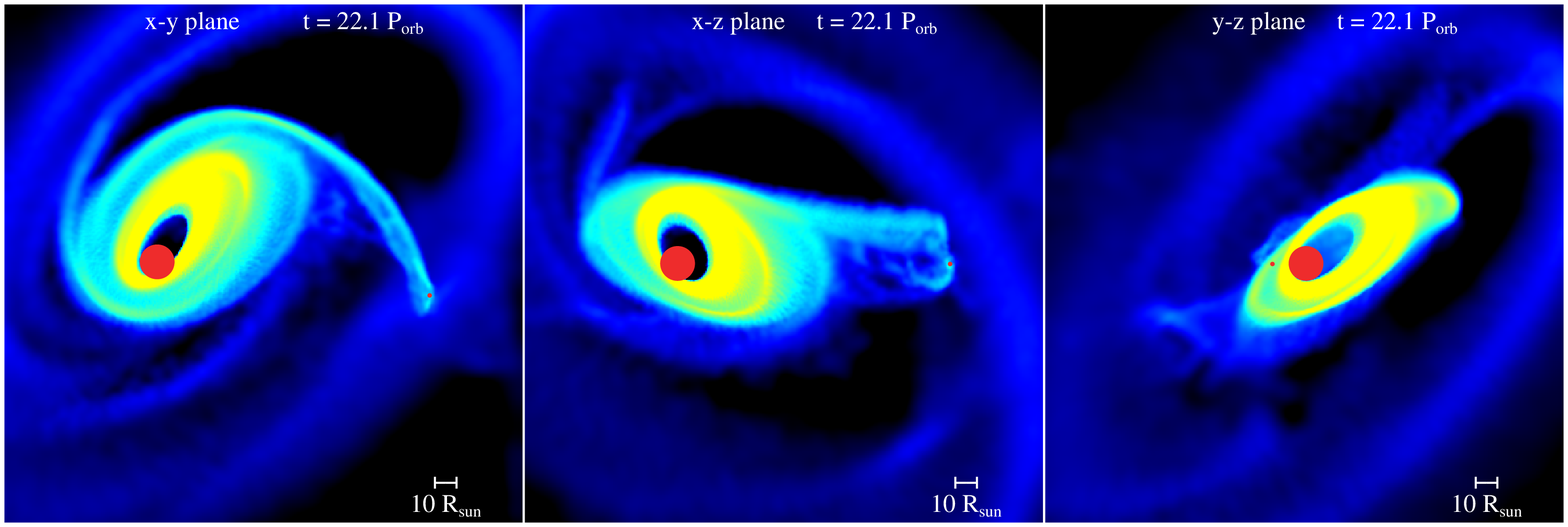}
\includegraphics[width=18cm]{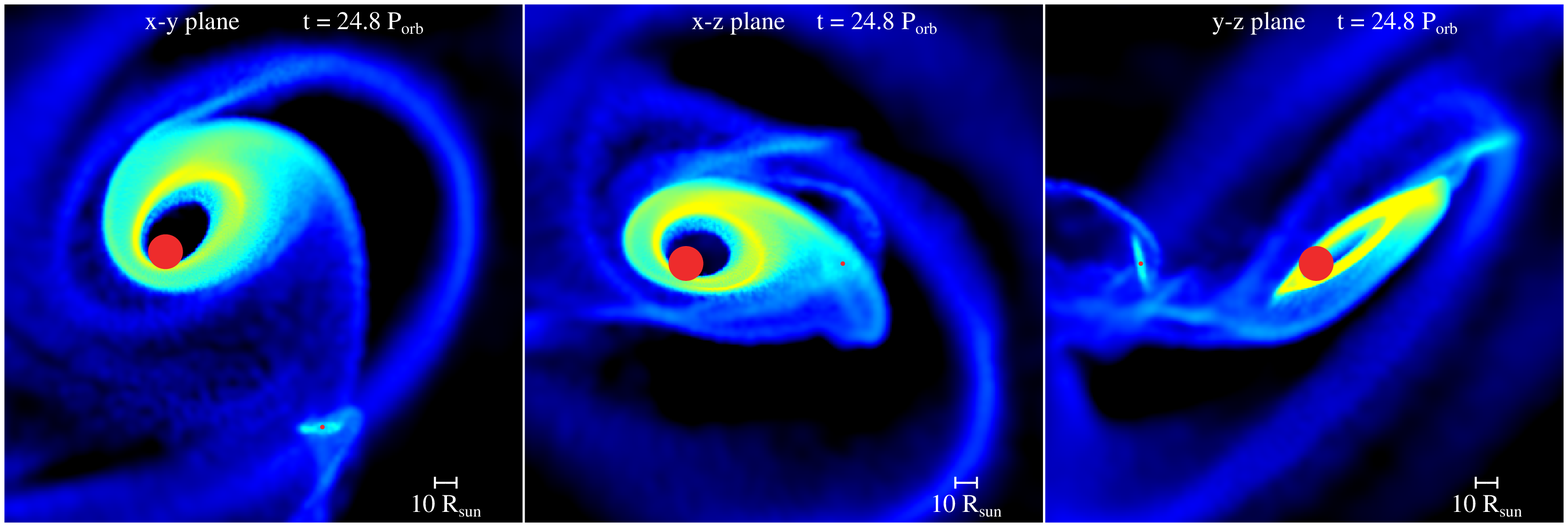}
\end{center}
\caption{Same as Fig.~\ref{plot1} but at times $t=22.1 \,P_{\rm. orb}$
  (upper panels) and $t=24.8\,P_{\rm. orb}$ (lower panels) during and
  after the Type~II outburst.  The Be star disk is eccentric. Top
  panels: The neutron star is passing through a spiral arm and
  accreting a high amount of mass in a Type~II outburst. Bottom
  panels: After the Type~II outburst a disk has formed around the
  neutron star. }
\label{plot3}
\end{figure*}

\begin{table}
\caption{Be/X-ray binary model properties } \centering
\begin{tabular}{lllll}
\hline
Be/X-ray Binary Parameters  \\
\hline
\hline
Mass of the Be star  & $M_\star/{\rm. M_\odot}$ & 18 \\
Radius of the Be star &  $R_\star/{\rm. R_\odot}$ & 8 \\
Mass of the neutron star & $M_{\rm. NS}/{\rm. M_\odot}$ & 1.4 \\
Mass of the binary & $M/{\rm. M_\odot}$ & 19.4 \\
\hline
Orbital Period & $P_{\rm. orb}/\rm. d$ & 24 \\
Semi-major axis & $a/{\rm. R_\odot}$ & 95 \\ 
Eccentricity & $e$ & 0.34 \\
\hline
Initial Be star disk mass & $M_{\rm. di}/{\rm. M_\odot}$ & $10^{-8}$ \\
Be star disk viscosity parameter & $\alpha$ & 0.2-0.4 \\
Be star disk aspect ratio & $H/R (R=R_\star)$ & 0.01 \\
Initial disk inclination & $i$ & $70^\circ$ \\ 
\hline
\end{tabular}
\label{tab}
\end{table}

Previous work on misaligned Be star systems mainly focused on coplanar
systems or those with a small misalignment angle, $i \lesssim
45^\circ$
\citep[e.g.][]{Okazaki2006,Okazakietal2007,Okazaki2007}. They found
that the resonant truncation radius for the Be star disk is the same
for the inclined case unless $i>60^\circ$ ($i$ is the angle between
the spin of the Be star and the binary orbital axis). In this work we
use the smoothed particle hydrodynamics (SPH) code {\sc. Phantom}
\citep{PF2010,LP2010} to model highly misaligned systems. {\sc
  Phantom} has been used to model misaligned disks in other binary
systems \citep{Nixon2012,Nixonetal2013} and we use the same method
here.  The simulations have two sink particles, one representing a Be
star with mass $M_\star=18\,\rm M_\odot$ and the other a neutron star
of mass $M_{\rm NS}=1.4\,\rm M_\odot$. The binary has an orbital
period of $P_{\rm orb}=24\,\rm d$ (thus the orbit has semi-major axis
$a=95\,\rm R_\odot$) and eccentricity $e=0.34$, to target the Be/X-ray
binary 4U 0115+63, although our results are more generally
applicable. The parameters of the simulation are summarized in
Table~\ref{tab}. Figs.~\ref{plot1}-\ref{plot3} show the overall
evolution of the calculation.

The accretion radius of the Be star is equal to its stellar radius of
$R_\star=8\,\rm R_\odot$.  For the neutron star we take the accretion
radius to be $1\,\rm R_\odot$ (the Bondi-Hoyle accretion radius at
periastron is about $3.3\,\rm R_\odot$).  Thus, we do not model the
inner parts of the accretion disk around the neutron star (if it
forms).  Fig.~\ref{plot1} shows the initial flat, but tilted disk with
an inclination of $i=70^\circ$ to the $x$-axis. The disk has a mass of
$M_{\rm di}=10^{-8}\,\rm M_\odot$ distributed with a surface density
power law $\Sigma \propto R^{-1}$ between $R_{\rm in}=R_\star$ and
$R_{\rm out}=50\,\rm R_\odot$ with $10^6$ SPH particles. Varying the
outer disk radius, we find little difference in the disk behaviour for
smaller disk radii once the disk has expanded to the tidal truncation
radius. The disk is isothermal with an aspect ratio $H/R=0.01$ at the
inner edge, where $R=R_\star$, and the \cite{SS1973} $\alpha$
parameter varies in the range $0.2-0.4$ \citep[e.g.][]{Kingetal2007}
over the disk (we implement the disk viscosity in the usual manner by
adapting the SPH artificial viscosity according to the procedure
described in \cite{LP2010}, using $\alpha_{\rm AV} = 6.0$ and
$\beta_{\rm AV} = 2.0$).  The Be star disc is resolved with
shell-averaged smoothing length per scale height
$\left<h\right>/H=0.5$.

\begin{figure*}
\includegraphics[width=8.4cm]{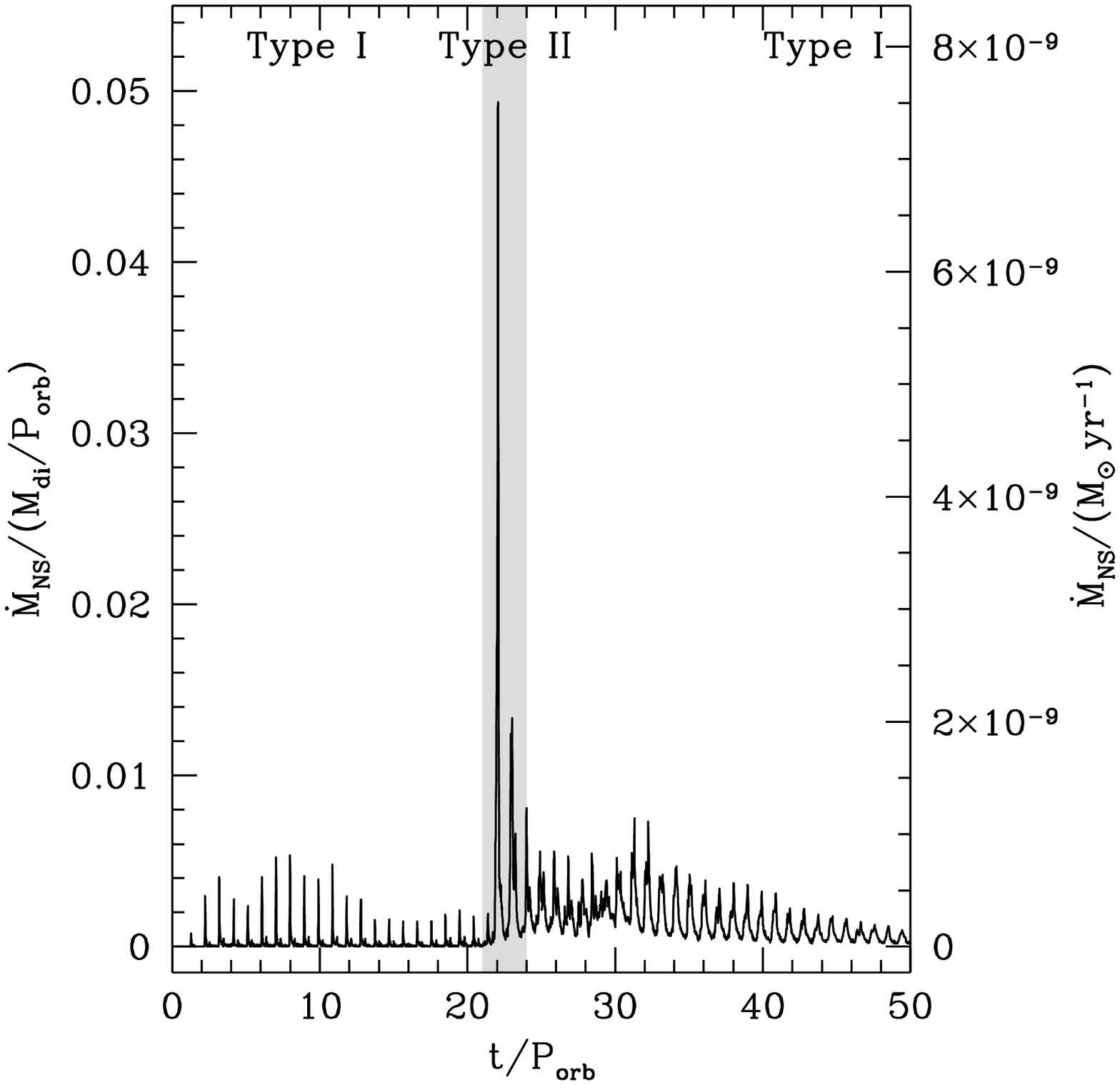}
\includegraphics[width=8.4cm]{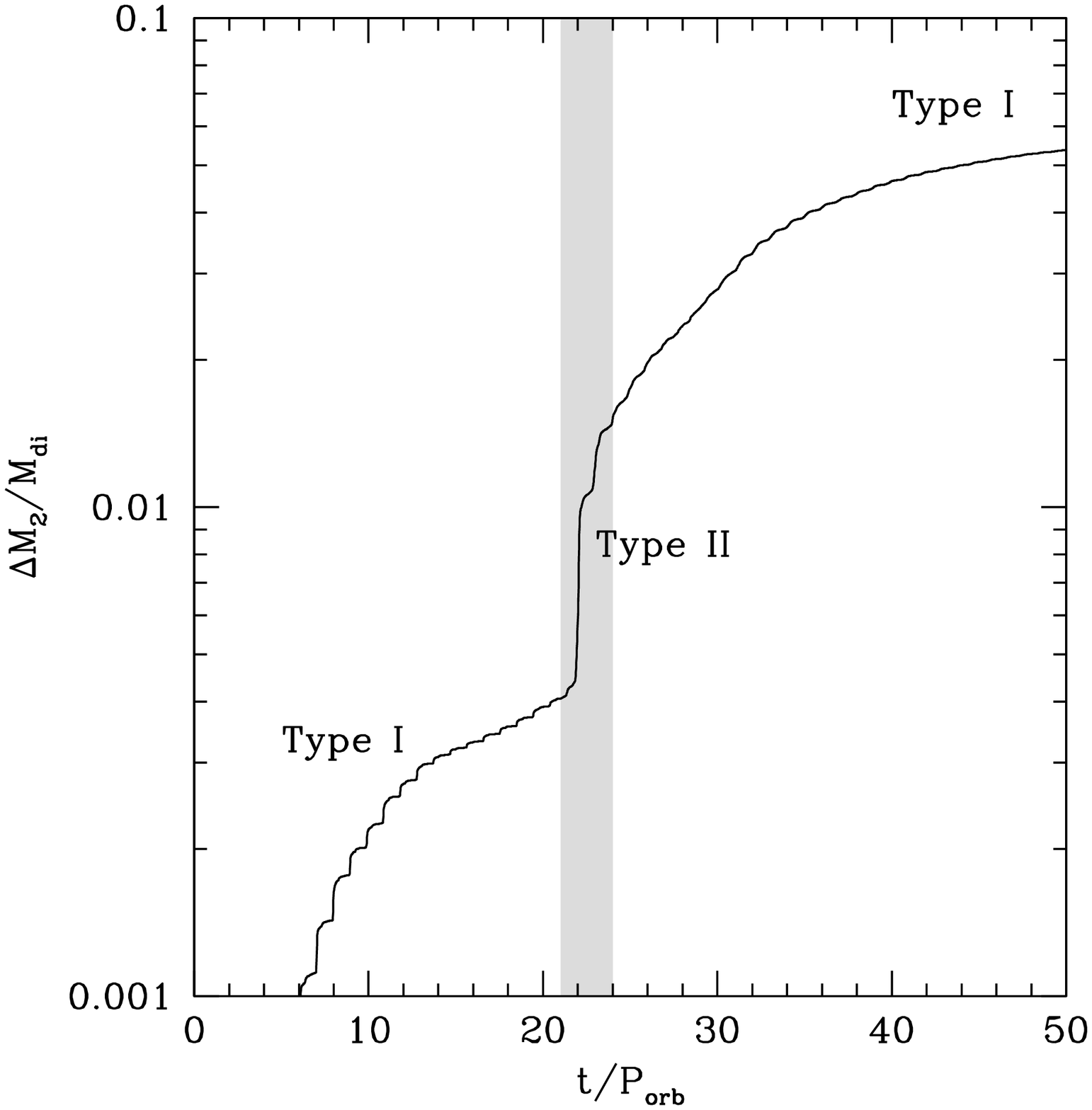}
\caption{Left: The mass transfer rate on to the neutron star (at the
  accretion radius of $1\,\rm R_\odot$). The right hand axis shows the
  equivalent accretion rate in a system with an initial Be star disk
  mass of $M_{\rm di}=10^{-8}\,\rm M_\odot$. Initially the system
  displays periodic Type~I accretion outbursts. From about a time of
  about 21 to $24\,P_{\rm orb}$, the system undergoes a Type~II
  outburst, where the average accretion rate on to the neutron star is
  significantly higher.  Right: The corresponding total mass accreted
  on to the neutron star. In each plot the shaded region show the
  approximate time of the Type~II outburst. }
\label{outburst}
\end{figure*}

The left hand panel of Fig.~\ref{outburst} shows the mass flow rate
through the neutron star accretion radius as a function of time.
Initially small (Type~I) accretion outbursts occur every orbital
period.  Fig.~\ref{plot2} show the disk structure during a Type~I
outburst at a time of $6.1\,P_{\rm orb}$. These outbursts occur when
the disk is close to circular and the neutron star passes close to the
spiral arm of the Be star disk. A small amount of material is pulled
from the spiral arm on to the neutron star.

Fig~\ref{outburst} shows that at a time of around $21\,P_{\rm orb}$
the accretion changes phase. The peak accretion rate is about an order
of magnitude higher than during the Type~I phase and the accretion
does not subside between the periastron passages. We suggest that this
is a giant Type~II outburst. This can be seen clearly in the right
panel of Fig.~\ref{outburst} where we show the total mass accreted on
to the neutron star as a function of time. The mass increases slowly
during the Type~I phase and then dramatically during the Type~II phase
before returning to the Type~I phase. The top row of panels in
Fig.~\ref{plot3} show the disk structure during the Type~II outburst
at a time of $22.1\,P_{\rm orb}$. The Be star disk is highly eccentric
and it is this that causes the Type~II outburst where a large amount
of material is transferred to the neutron star. The bottom row of
panels show the end of the outburst and the neutron star disk. The
precession of the eccentric Be star disk can also be seen by comparing
the disk structures at different times.

We bin the particles in (instantaneous) radius from the Be star and
calculate properties of the Be star disk. Initially the disk is
misaligned by $i=70^\circ$. Just before the Type~II outburst, the disk
has aligned to $i\approx 60^\circ$.  The disk remains approximately
flat during this process with the outer parts aligning only a few
degrees more than the inner parts. This is in agreement with the
analytic estimates for the alignment timescale for the Be star disk
with the binary orbit of the order of a few years
\citep{Martinetal2011}.  Furthermore, we find that the eccentricity
grows first at large radius in the disk. Just before the Type~II
outburst, the eccentricity of the disk is approximately constant with
radius at $e=0.6$. During and after the giant outburst, the
eccentricity decays.

\subsection{Eccentricity Growth in the Be star disk}

There are several mechanisms that may drive eccentricity in a
circumprimary disk that we discuss here.  First, eccentricity may be
directly excited due to the eccentric companion. A perturber on an
eccentric orbit forces an eccentricity on an initially circular
particle orbit. The particle eccentricity grows nearly linearly in
time (see Section III.c of \cite{Goldreich1980}).  We have performed
test particle simulations of a coplanar eccentric binary system
including the eccentric forcing of the companion \citep[see
  equation~(9) in][]{Goldreich1980}. The initially constant rate of
eccentricity growth of the particle increases strongly with distance
from the Be star, $d$, as $de/dt \propto d^{2.8}$ (for $d \lesssim
0.55\,a$). For example, a particle with an initial distance of
$d=0.5\,\rm a$, the eccentricity grows linearly at a rate of $de/dt
\approx 0.05$ per binary orbit.  Thus, the disk becomes eccentric on a
timescale of around 10 orbital periods. The eccentricity of the
particle orbit grows perpendicularly to the eccentricity of the binary
orbit. This is observed in our simulations and thus we suggest that
gravitational forcing from the companion is the most likely mechanism
for eccentricity growth here.  Hence, the disk must expand radially in
order for eccentricity to be driven. We will investigate this process
in the inclined disk in more detail in a future publication.

Secondly, instability at the 3:1 Lindblad resonance can lead to
eccentricity growth even in an initially circular disk
\citep{Lubow1991,Lubow1991b,Lubow1992}. The radius of the 3:1
resonance is $R_{\rm res}=3^{-2/3}(1+q)^{-1/3}a$
\citep[e.g.][]{Goodchild2006}, which for our typical parameters is
$R_{\rm res}\approx 44.5\,\rm R_\odot$. Finally, it could be possible
for eccentricity to be driven by the stream from the outside of the Be
star disk falling back towards and hitting the Be star disk
\citep{Lubow1994,Shi2012}. However, there is only a small amount
of material involved in the streams.

In the coplanar case, the tidal truncation radius of the disk may be
estimated by the radius at which ballistic particle orbits cross. In a
circular binary with our parameters this would be at a radius of
$0.48a$ \citep{Paczynski1977}. For our parameters this is
$45.6(a/95{\,\rm R_\odot})\,\rm R_\odot$.  In an eccentric binary, the
size of the disk is further reduced. For our parameters, the
truncation radius is around $0.38\,a$ \citep{Artymowicz1994}. In an
eccentric binary, the disk eccentricity (with no dissipation) grows at
all radii but the rate is strongly dependent on radius. Thus large
eccentricity growth is limited to high inclination disks where reduced
tidal torques allow the disk to expand to large radii (Nixon et al. in
prep). For smaller disks (lower inclinations), dissipation can keep
the disk close to circular with $e\lesssim 0.1$
\citep{Okazaki2002}. We have run several SPH simulations with varying
misalignment angle and we find that the disk becomes significantly
eccentric only for $i \gtrsim 60^\circ$.  Thus, Type~II outbursts are
expected only in systems with a large misalignment angle.

\section{Accretion disk around the Neutron Star}
\label{NS}

Here we consider properties of the accretion disk that forms around
the neutron star. Due to the misalignment between the Be star disk and
the neutron star orbit it is likely that accretion on to the neutron
star proceeds through an initially misaligned disk
\citep[c.f.][]{NixonandSalvessen2014}.  For numerical tractability we
cannot model the neutron star disk in its entirety and it remains
under-resolved even during the peak of the Type~II outburst. This will
be investigated in more detail in a future publication.  However, we
have run several SPH simulations with a much smaller neutron star
accretion radius. We do not find significant evidence for a neutron
star accretion disk during the Type~I activity, only during the
Type~II. There is much less material and its circularization radius is
much smaller during the Type~I outburst phase and thus it accretes
more rapidly \citep[see also][]{Okazaki2002}.

The viscous timescale in the neutron star disk is $\tau_{\nu_{\rm
    NS}}=R_{\rm NS}^2/\nu_{\rm NS}$, where $R_{\rm NS}$ is the
distance from the neutron star, the viscosity is $\nu_{\rm
  NS}=\alpha_{\rm NS} c_{\rm s,NS}^2/\Omega_{\rm NS}$, $\alpha_{\rm
  NS}$ is the \cite{SS1973} parameter and the sound speed is $c_{\rm
  s,NS}=\sqrt{{\cal R}T/\mu}$ where $\cal R$ is the gas constant and
$\mu$ is the gas mean molecular weight. The temperature of the disk
surface is found by assuming a steady profile
\begin{equation}
\sigma T^4=\frac{3 \dot M_{\rm NS}}{8 \pi}\Omega_{\rm NS}^2
\end{equation}
\citep{Pringle1981}.  Thus, the accretion timescale of the disk scaled
by the orbital period is
\begin{equation}
\frac{\tau_{\nu_{\rm NS}}}{P_{\rm orb}} =4.7  \, \left(\frac{\alpha_{\rm NS}}{0.3}\right)^{-1}
\left(\frac{M_{\rm NS}}{1.4\,\rm M_\odot}\right)^{\frac{1}{4}}
\left(\frac{\dot M_{\rm NS}}{10^{-8}\,\rm
  M_\odot\,yr^{-1}}\right)^{-\frac{1}{4}} \left(\frac{R_{\rm NS}}{0.5\,\rm
  R_\odot}\right)^\frac{3}{4}.
\end{equation}
If the disc is significantly optically thick this is an overestimate.
While a radially extended neutron star disk is present, the accretion
on to the neutron star is smoothed on a timescale of several orbital
periods. This explains why there is no orbital modulation observed
during a Type~II X-ray outburst. We do not expect smoothing on this
timescale to take place during the Type~I activity because the
accreting material has lower initial angular momentum and therefore a
smaller circularisation radius ($\lesssim 0.5\,\rm R_\odot$) and
viscous time. These numbers are approximate and tend to over-predict
the viscous time compared to observations as we assume a steady
disk. Time dependent processes, such as tides or interactions of the
neutron star disk with streams from the Be star disk could shorten the
viscous timescale.

The corresponding X-ray luminosity of the disk is $L_{\rm X}=GM_{\rm NS} \dot
M_{\rm NS}/r_{\rm NS}$, where $r_{\rm NS}$ is the radius of the neutron star
and thus
\begin{equation}
L_{\rm X}=1.2\times 10^{38}
\left(\frac{M_{\rm NS}}{1.4\,\rm  M_\odot}\right)\left(\frac{\dot M_{\rm NS}}{10^{-8}\rm
  M_\odot\,yr^{-1}}\right)\left(\frac{r_{\rm NS}}{10^6\,\rm cm}\right)^{-1}\,\rm erg\,s^{-1}.
\end{equation}
Thus, for the accretion rates to agree with observations of the X-ray
luminosities, we require an initial disk mass of around $10^{-8}\,\rm
M_\odot$ or smaller. This agrees well with the Be star mass loss rates
that are in the range $10^{-8}$ to $10^{-10}\,\rm M_\odot\,yr^{-1}$
\citep[e.g.][]{Snow1981,Porter2003} and the lifetimes of the Be star
disks (of a few years).

\section{discussion}
\label{discussion}

The simulations presented here model a warped, eccentric, precessing
Be star disk.  The ejection mechanism for material from the Be star to
the disk is poorly understood and thus we have considered only the
evolution following disk formation. Thus, the disc is an accretion
rather than a decretion disc, but over the timescales we simulate this
should not affect our conclusions. The ejection of material from the
Be Star surface in a circular orbit at the inner edge of the disc
could affect the eccentricity growth and inclination of the
disc. However, because eccentricity growth is driven more strongly in
the outer parts of the disc, this should not affect the Type~II
outburst mechanism.

During the Type~II outburst in our simulation, the Be star disk
becomes significantly depleted. About $40\%$ of the initial disk mass
is accreted on to the Be star. This depletion could explain the cycles
observed in the optical light curves of Be/X-ray binaries
\citep[e.g.][]{Charles2011}. With mass ejection from the Be star
included, the disk would build up again and we would expect the
Type~II outbursts to recur. The timescale of recurrence would depend
on the accretion rate on to the Be star disk.  Parameters such as the
eccentricity, misalignment and binary separation could be varied to
model other Be/X-ray binary systems and predict the types of outbursts
that may occur. 

Observations suggest that the Type~II outbursts occur when the disk
becomes warped. However we have found that this is not a sufficient
condition, it is necessary for the disk to become eccentric. An
inclined circular disk, which can expand freely as the tidal torques
are smaller, becomes eccentric before it becomes large enough for a NS
encounter.  If there are no outbursts observed in a Be binary system,
the binary must have small eccentricity and misalignment. Conversely,
if Type~II outbursts are observed, then the system must be highly
misaligned.  We find that eccentricity may also be driven in counter
rotating discs. The frequency of outbursts in Be/X-ray binary systems
will be investigated in future with a disc model including mass
injection into the disc from the Be~star in order to determine
critical inclination angles required for outbursts.  The tidal torque
aligns the disk with the binary orbital plane on a timescale of years
\citep{Martinetal2011}. In a long term disk simulation this alignment
may also play a role in ending the outburst cycle.

\section{Conclusions}

We have performed 3D SPH simulations of a moderately eccentric and
highly misaligned Be/X-ray binary. We find that for a binary with
orbital period of $24\,\rm d$ and eccentricity $0.34$ that Type I
outbursts occur for all misalignments, including the aligned case. If
the disk is sufficiently misaligned, $i \gtrsim 60^\circ$, the disk is
much larger because the tidal torque is weaker. This allows strong
eccentricity growth in the outer parts of the disk from the
gravitational forcing of the eccentric companion. Thus, the disk
becomes eccentric on a timescale of around ten binary orbital
periods. We suggest that the eccentric disk causes a Type~II X-ray
outburst as a large amount of material is transferred to the neutron
star.

\section*{Acknowledgments} 
RGM's support was provided under contract with the California
Institute of Technology (Caltech) funded by NASA through the Sagan
Fellowship Program.  Support for CJN was provided by NASA through the
Einstein Fellowship Program, grant PF2-130098. PJA acknowledges
support from NASA's ATP program under awards~NNX11AE12G and
NNX14AB42G. DJP is supported by Future Fellowship FT130100034 from the
Australian Research Council. We acknowledge the use of SPLASH
\citep{Price2007} for the rendering of the figures. This work utilised
the Janus supercomputer, which is supported by the National Science
Foundation (award number CNS-0821794), the University of Colorado
Boulder, the University of Colorado Denver, and the National Center
for Atmospheric Research. The Janus supercomputer is operated by the
University of Colorado Boulder.

\bibliographystyle{apj}

\begin{thebibliography}{50}
\expandafter\ifx\csname natexlab\endcsname\relax\def\natexlab#1{#1}\fi

\bibitem[{{Artymowicz} \& {Lubow}(1994)}]{Artymowicz1994}
{Artymowicz}, P. \& {Lubow}, S.~H. 1994, \apj, 421, 651

\bibitem[{{Bjorkman} {et~al.}(2002){Bjorkman}, {Miroshnichenko}, {McDavid}, \&
  {Pogrosheva}}]{Bjorkman2002}
{Bjorkman}, K.~S., {Miroshnichenko}, A.~S., {McDavid}, D., \& {Pogrosheva},
  T.~M. 2002, \apj, 573, 812

\bibitem[{{Carciofi}(2011)}]{Carciofi2011}
{Carciofi}, A.~C. 2011, in IAU Symposium, Vol. 272, IAU Symposium, ed.
  C.~{Neiner}, G.~{Wade}, G.~{Meynet}, \& G.~{Peters}, 325--336

\bibitem[{{Goldreich} \& {Tremaine}(1980)}]{Goldreich1980}
{Goldreich}, P. \& {Tremaine}, S. 1980, \apj, 241, 425

\bibitem[{{Goodchild} \& {Ogilvie}(2006)}]{Goodchild2006}
{Goodchild}, S. \& {Ogilvie}, G. 2006, \mnras, 368, 1123

\bibitem[{{Hanuschik}(1996)}]{Hanuschik1996}
{Hanuschik}, R.~W. 1996, \aap, 308, 170

\bibitem[{{Kato}(2014)}]{Kato2014}
{Kato}, S. 2014, ArXiv e-prints

\bibitem[{{King} {et~al.}(2007){King}, {Pringle}, \& {Livio}}]{Kingetal2007}
{King}, A.~R., {Pringle}, J.~E., \& {Livio}, M. 2007, \mnras, 376, 1740

\bibitem[{{Kretschmar} {et~al.}(2013){Kretschmar}, {Nespoli}, {Reig}, \&
  {Anders}}]{Kretschmar2013}
{Kretschmar}, P., {Nespoli}, E., {Reig}, P., \& {Anders}, F. 2013, ArXiv
  e-prints

\bibitem[{{Lee} {et~al.}(1991){Lee}, {Osaki}, \& {Saio}}]{Lee1991}
{Lee}, U., {Osaki}, Y., \& {Saio}, H. 1991, \mnras, 250, 432

\bibitem[{{Lodato} \& {Price}(2010)}]{LP2010}
{Lodato}, G. \& {Price}, D.~J. 2010, \mnras, 405, 1212

\bibitem[{{Lubow}(1991{\natexlab{a}})}]{Lubow1991}
{Lubow}, S.~H. 1991{\natexlab{a}}, \apj, 381, 259

\bibitem[{{Lubow}(1991{\natexlab{b}})}]{Lubow1991b}
---. 1991{\natexlab{b}}, \apj, 381, 268

\bibitem[{{Lubow}(1992)}]{Lubow1992}
---. 1992, \apj, 401, 317

\bibitem[{{Lubow}(1994)}]{Lubow1994}
---. 1994, \apj, 432, 224

\bibitem[{{Martin} {et~al.}(2011){Martin}, {Pringle}, {Tout}, \&
  {Lubow}}]{Martinetal2011}
{Martin}, R.~G., {Pringle}, J.~E., {Tout}, C.~A., \& {Lubow}, S.~H. 2011,
  \mnras, 416, 2827

\bibitem[{{Martin} {et~al.}(2009){Martin}, {Tout}, \&
  {Pringle}}]{Martinetal2009b}
{Martin}, R.~G., {Tout}, C.~A., \& {Pringle}, J.~E. 2009, \mnras, 397, 1563

\bibitem[{{Martin} {et~al.}(2010){Martin}, {Tout}, \&
  {Pringle}}]{Martinetal2010}
---. 2010, \mnras, 401, 1514

\bibitem[{{Moritani} {et~al.}(2011){Moritani}, {Nogami}, {Okazaki}, {Imada},
  {Kambe}, {Honda}, {Hashimoto}, \& {Ichikawa}}]{Moritanietal2011}
{Moritani}, Y., {Nogami}, D., {Okazaki}, A.~T., {Imada}, A., {Kambe}, E.,
  {Honda}, S., {Hashimoto}, O., \& {Ichikawa}, K. 2011, \pasj, 63, L25

\bibitem[{{Moritani} {et~al.}(2013){Moritani}, {Nogami}, {Okazaki}, {Imada},
  {Kambe}, {Honda}, {Hashimoto}, {Mizoguchi}, {Kanda}, {Sadakane}, \&
  {Ichikawa}}]{Moritani2013}
{Moritani}, Y., {Nogami}, D., {Okazaki}, A.~T., {Imada}, A., {Kambe}, E.,
  {Honda}, S., {Hashimoto}, O., {Mizoguchi}, S., {Kanda}, Y., {Sadakane}, K.,
  \& {Ichikawa}, K. 2013, \pasj, 65, 83

\bibitem[{{Munar-Adrover} {et~al.}(2014){Munar-Adrover}, {Paredes}, {Rib{\'o}},
  {Iwasawa}, {Zabalza}, \& {Casares}}]{MunarAdrover2014}
{Munar-Adrover}, P., {Paredes}, J.~M., {Rib{\'o}}, M., {Iwasawa}, K.,
  {Zabalza}, V., \& {Casares}, J. 2014, ArXiv e-prints

\bibitem[{{Negueruela} {et~al.}(2001){Negueruela}, {Okazaki}, {Fabregat},
  {Coe}, {Munari}, \& {Tomov}}]{Negueruela2001}
{Negueruela}, I., {Okazaki}, A.~T., {Fabregat}, J., {Coe}, M.~J., {Munari}, U.,
  \& {Tomov}, T. 2001, \aap, 369, 117

\bibitem[{{Negueruela} {et~al.}(1998){Negueruela}, {Reig}, {Coe}, \&
  {Fabregat}}]{Negueruela1998}
{Negueruela}, I., {Reig}, P., {Coe}, M.~J., \& {Fabregat}, J. 1998, \aap, 336,
  251

\bibitem[{{Nixon} {et~al.}(2013){Nixon}, {King}, \& {Price}}]{Nixonetal2013}
{Nixon}, C., {King}, A., \& {Price}, D. 2013, \mnras, 434, 1946

\bibitem[{{Nixon} \& {Salvesen}(2014)}]{NixonandSalvessen2014}
{Nixon}, C. \& {Salvesen}, G. 2014, \mnras, 437, 3994

\bibitem[{{Nixon}(2012)}]{Nixon2012}
{Nixon}, C.~J. 2012, \mnras, 423, 2597

\bibitem[{{Okazaki}(2006)}]{Okazaki2006}
{Okazaki}, A.~T. 2006, Boletin de la Asociacion Argentina de Astronomia La
  Plata Argentina, 49, 321

\bibitem[{{Okazaki}(2007)}]{Okazaki2007}
{Okazaki}, A.~T. 2007, in Astronomical Society of the Pacific Conference
  Series, Vol. 367, Massive Stars in Interactive Binaries, ed. N.~{St.-Louis}
  \& A.~F.~J. {Moffat}, 485

\bibitem[{{Okazaki} {et~al.}(2002){Okazaki}, {Bate}, {Ogilvie}, \&
  {Pringle}}]{Okazaki2002}
{Okazaki}, A.~T., {Bate}, M.~R., {Ogilvie}, G.~I., \& {Pringle}, J.~E. 2002,
  \mnras, 337, 967

\bibitem[{{Okazaki} \& {Hayasaki}(2007)}]{Okazakietal2007}
{Okazaki}, A.~T. \& {Hayasaki}, K. 2007, in Astronomical Society of the Pacific
  Conference Series, Vol. 361, Active OB-Stars: Laboratories for Stellare and
  Circumstellar Physics, ed. A.~T. {Okazaki}, S.~P. {Owocki}, \& S.~{Stefl},
  395

\bibitem[{{Okazaki} {et~al.}(2013){Okazaki}, {Hayasaki}, \&
  {Moritani}}]{Okazaki2013}
{Okazaki}, A.~T., {Hayasaki}, K., \& {Moritani}, Y. 2013, \pasj, 65, 41

\bibitem[{{Okazaki} \& {Negueruela}(2001)}]{Okazaki2001}
{Okazaki}, A.~T. \& {Negueruela}, I. 2001, \aap, 377, 161

\bibitem[{{Paczynski}(1977)}]{Paczynski1977}
{Paczynski}, B. 1977, \apj, 216, 822

\bibitem[{{Porter}(1996)}]{Porter1996}
{Porter}, J.~M. 1996, \mnras, 280, L31

\bibitem[{{Porter} \& {Rivinius}(2003)}]{Porter2003}
{Porter}, J.~M. \& {Rivinius}, T. 2003, \pasp, 115, 1153

\bibitem[{{Price}(2007)}]{Price2007}
{Price}, D.~J. 2007, Pasa, 24, 159

\bibitem[{{Price} \& {Federrath}(2010)}]{PF2010}
{Price}, D.~J. \& {Federrath}, C. 2010, \mnras, 406, 1659

\bibitem[{{Pringle}(1981)}]{Pringle1981}
{Pringle}, J.~E. 1981, \araa, 19, 137

\bibitem[{{Pringle}(1991)}]{Pringle1991}
---. 1991, \mnras, 248, 754

\bibitem[{{Rajoelimanana} {et~al.}(2011){Rajoelimanana}, {Charles}, \&
  {Udalski}}]{Charles2011}
{Rajoelimanana}, A.~F., {Charles}, P.~A., \& {Udalski}, A. 2011, \mnras, 413,
  1600

\bibitem[{{Reig}(2011)}]{Reig2011}
{Reig}, P. 2011, \apss, 332, 1

\bibitem[{{Reig} {et~al.}(2007){Reig}, {Larionov}, {Negueruela}, {Arkharov}, \&
  {Kudryavtseva}}]{Reigetal2007}
{Reig}, P., {Larionov}, V., {Negueruela}, I., {Arkharov}, A.~A., \&
  {Kudryavtseva}, N.~A. 2007, \aap, 462, 1081

\bibitem[{{Rivinius} {et~al.}(2001){Rivinius}, {Baade}, {{\v S}tefl},
  {Townsend}, {Stahl}, {Wolf}, \& {Kaufer}}]{Rivinius2001}
{Rivinius}, T., {Baade}, D., {{\v S}tefl}, S., {Townsend}, R.~H.~D., {Stahl},
  O., {Wolf}, B., \& {Kaufer}, A. 2001, \aap, 369, 1058

\bibitem[{{Shakura} \& {Sunyaev}(1973)}]{SS1973}
{Shakura}, N.~I. \& {Sunyaev}, R.~A. 1973, \aap, 24, 337

\bibitem[{{Shi} {et~al.}(2012){Shi}, {Krolik}, {Lubow}, \& {Hawley}}]{Shi2012}
{Shi}, J.-M., {Krolik}, J.~H., {Lubow}, S.~H., \& {Hawley}, J.~F. 2012, \apj,
  749, 118

\bibitem[{{Slettebak}(1982)}]{Slettebak1982}
{Slettebak}, A. 1982, \apjs, 50, 55

\bibitem[{{Smith} {et~al.}(1994){Smith}, {Hubeny}, {Lanz}, \&
  {Meylan}}]{Smith1994}
{Smith}, M.~A., {Hubeny}, I., {Lanz}, T., \& {Meylan}, T. 1994, \apj, 432, 392

\bibitem[{{Snow}(1981)}]{Snow1981}
{Snow}, Jr., T.~P. 1981, \apj, 251, 139

\bibitem[{{Stella} {et~al.}(1986){Stella}, {White}, \& {Rosner}}]{Stella1986}
{Stella}, L., {White}, N.~E., \& {Rosner}, R. 1986, \apj, 308, 669

\bibitem[{{Underhill}(1987)}]{Underhill1987}
{Underhill}, A.~B. 1987, in IAU Colloq. 92: Physics of Be Stars, ed.
  A.~{Slettebak} \& T.~P. {Snow}, 411--425

\end{thebibliography}

\label{lastpage}
\end{document}